\documentstyle[draft,epsf]{article}

\def\be{\begin{equation}}
\def\bea{\begin{eqnarray}}
\def\bma{\begin{mathletters}}
\def\ee{\end{equation}}
\def\eea{\end{eqnarray}}
\def\ema{\end{mathletters}}

\begin{document}
\author{Vlatko Vedral \footnote {Permanent address: Optics Section, Blackett Laboratory, Imperial College
London SW7 2BZ, England; e-mail: v.vedral@ic.ac.uk}}
\title{Geometric Phases and Topological Quantum Computation \footnote{Based on the set of lectures delivered in
October $2002$ at the International Centre for Theoretical Physics
in Trieste.}}
\date{\today}

\maketitle

\begin{abstract}
In the first part of this review we introduce the basics theory
behind geometric phases and emphasize their importance in quantum
theory. The subject is presented in a general way so as to
illustrate its wide applicability, but we also introduce a number
of examples that will help the reader understand the basic issues
involved. In the second part we show how to perform a universal
quantum computation using only geometric effects appearing in
quantum phases. It is then finally discussed how this geometric
way of performing quantum gates can lead to a stable, large scale,
intrinsically fault-tolerant quantum computer.
\end{abstract}

\section{\protect\bigskip Introduction}

This review is on the nature of geometric phases in quantum
mechanics and their use in quantum computation. I will introduce
the notion of a geometric phase in a general way which is neither
traditional nor historical. This will, I believe, help us
appreciate the importance of this notion and also its generality -
it is a phenomenon that has been discovered and exploited to
explain a number of effects in many different areas of physics
(see the collection of papers in \cite{Wilczek}). One of the
potential applications of this phase which is of interest to us is
in quantum computation (the standard text is \cite{Nielsen}). By
using geometric phases to implement any elementary quantum gate we
can therefore perform any quantum computation. The main reason why
we would want to use geometrical phases for quantum computation is
that frequently geometrical evolutions are easier to control and
may also be more resistant to random noise coming from the
environment. We will see, in Section $4$, how any quantum
computation can be implemented by using purely geometric
(topological) phases. This review is nether complete nor detailed
- I have tried to emphasize points and connections that are
usually neglected in literature, and to arrive at a potentially
very important modern application in quantum computing. I have
left a number of exercises and problems for the readers to address
on their own. My intention is for the reader to acquire the basic
flavor of the field which will then provide a good basis for a
more independent study and work.

The review is organized as follows: In section $2$, I will review
the notion of the geometric phase for the gauge invariant
perspective. This section will also present a method of measuring
geometric phases and discuss the classical analogue of the
geometric phase. Various links with differential geometry will be
emphasized. In section $3$, I review some basic notions in quantum
computation and present some simple algorithms. This is a
necessary background to be able to show why geometrical phases can
be used to implement any quantum computation. Section $4$ then
presents the most general way of implementing geometric quantum
evolution, via the so called non-Abelian geometric phases. An
example is given of a system that can support such an
implementation. Section $5$, finally, summarizes the review and
offers an outlook on the field.

\section{Geometric Phases}

Quantum states are represented as vectors in a complex vector
space (with a few other properties we need not worry about here).
However, these vectors are only defined up to a unit modulus
complex number, or a phase in other words. So, the two states
$|\xi\rangle$ and $e^{i\mu} |\xi\rangle$ are indistinguishable as
far as quantum mechanics is concerned - no set of measurements can
discriminate them. It is interesting that although this statement
looks fairly innocent at first sight, it can lead to some very
profound conclusions. The route we will take here and that turns
out to be very fruitful is to think of this extra phase in a
geometric way as well - where we represent the phase as a unit
length vector in the complex plane diagram.

\begin{figure}[ht]
\begin{center}
\hspace{0mm} \epsfxsize=7.0cm
\epsfbox{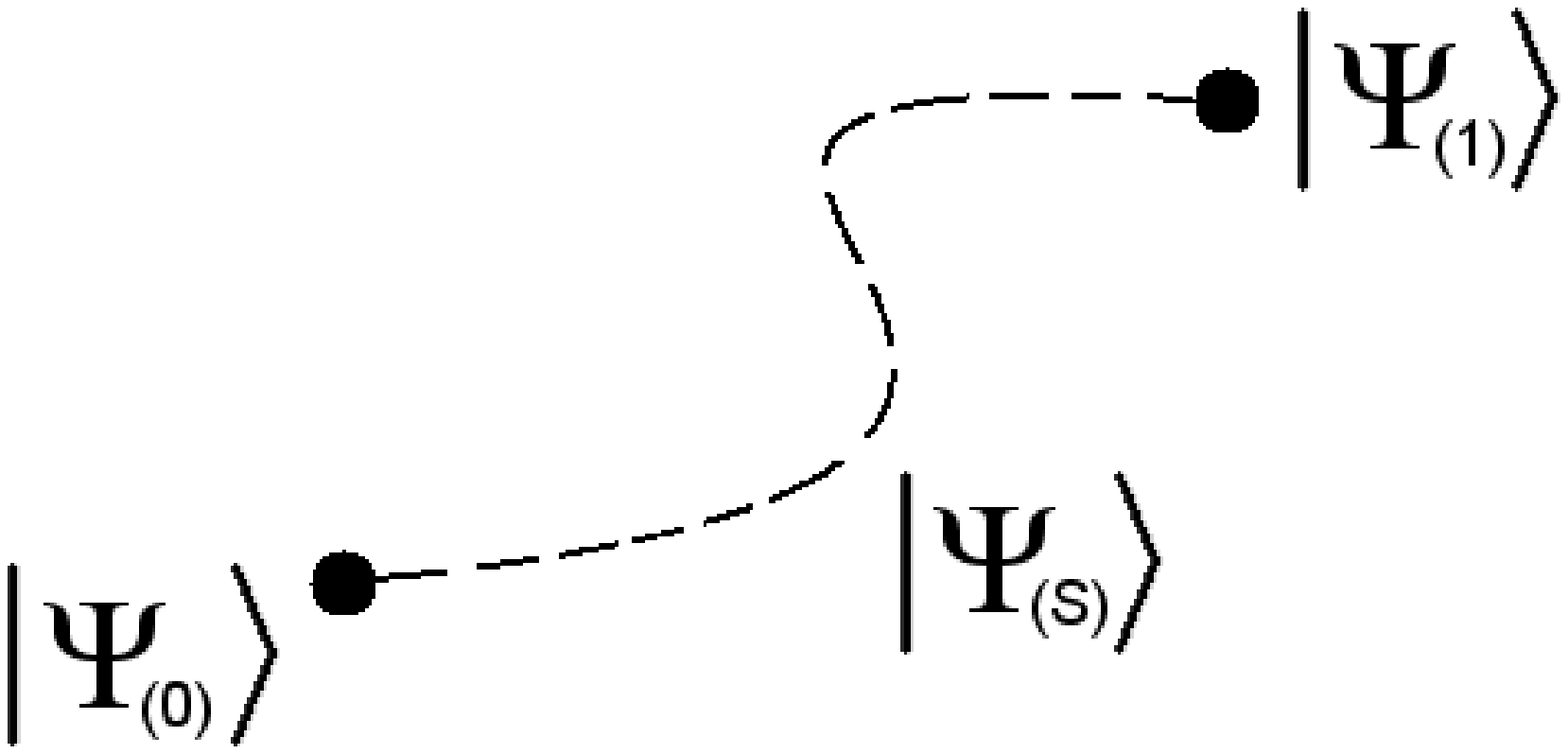}\\[0.2cm]
\begin{caption}
{How do we compare phases of two different states when the
probability between the states is the only experimentally well
defined quantity in quantum mechanics?}
\end{caption}
\end{center}
\end{figure}

Although we think of the amplitudes between two quantum states as
fundamental entities in quantum theory, it is only the
corresponding probabilities that we can ever observe
experimentally. Therefore, given two states $|\psi_i\rangle$ and
$|\psi_f\rangle$, we can only measure the mod square of their
overlap, $|\langle \psi_i|\psi_f\rangle|^2$. What happens then if
we wish to know the relative phase between these two states? Can
we define their relative phase in spite of the fact that the
absolute phase has no observable consequences? One idea to do so
may be to look at the amplitude between the two states in the
polar decomposition:
\begin{equation}
\langle \psi_i|\psi_f\rangle = re^{i\theta_{if}}
\end{equation}
and define the phase $\theta_{if}$ as the relative phase between
the two states. The main problem with this definition is that the
states $e^{i\alpha}|\psi_i\rangle$ and $e^{i\beta} |\psi_f\rangle$
which differ from the original states by an overall arbitrary
phase will have a different relative phases by the amount of
$\Delta \theta = \alpha - \beta$. This is not very satisfactory as
there are infinitely many choices for $\Delta \theta$ and they all
look equally appropriate. We formally say that this definition of
phase is {\em gauge dependent} (meaning ``phase dependent") - a
feature that is considered negative as all observable quantities
should be independent of the choice of the gauge. How about
defining a path connecting the two states, $|\psi (s) \rangle$,
such that when $s=0$ we have $|\psi_i\rangle$ and when $s=1$ we
have $|\psi_f\rangle$ (see Fig 1)? Then we can transport the
states $|\psi (s) \rangle$ from the position $i$ to the position
$f$ and see how different the final phase is to that of
$|\psi_f\rangle$ through {\em interference}. And if the states
$|\psi_f\rangle$ and $|\psi_i\rangle$ transported to
$|\psi_f\rangle$ interfere constructively, then they are in phase
and if they interfere destructively, they are out of phase; the
degree of interference can therefore define the phase difference
(I will be as vague as possible about what kind of interference
this is - I will be much more precise later on in subsection
$2,5$). But which path do we choose and how do we know that the
transport itself doesn't introduce any additional ``twists and
turns" in the phases so that we are actually comparing some
different phases to the original ones? This, in fact, is a well
known problem in (differential) geometry (see \cite{Wheeler} for
an excellent introduction). Let's state the problem in an abstract
setting first (those of you familiar with general relativity will
know immediately what I am taking about) and then specialize to
quantum geometric phases.

\subsection{Parallel Transport}

\begin{figure}[ht]
\begin{center}
\hspace{0mm} \epsfxsize=7.0cm
\epsfbox{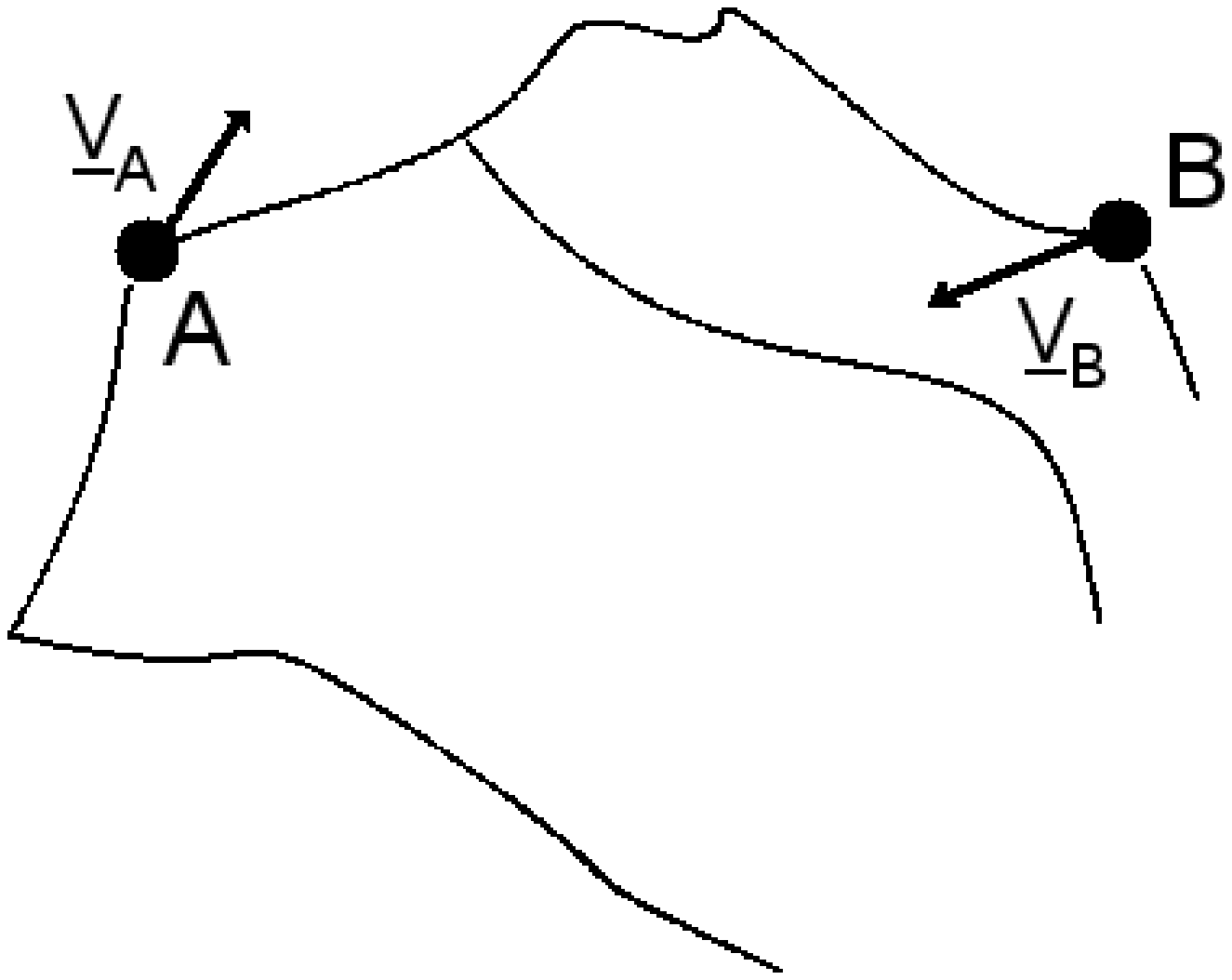}\\[0.2cm]
\begin{caption}
{Comparing direction of two vectors at two different points is a
non-trivial problem on a curved manifold. We have to somehow
transport the vectors to the same point and then see what the
angle between them is. Otherwise, comparing two vectors at
different locations is difficult.}
\end{caption}
\end{center}
\end{figure}

Suppose we have a curved manifold (as in Fig. 2) and we have a
vector at a point $A$ and another at a point $B$. What is the
angle between the two vectors (since phases in quantum mechanics
can be thought of as vectors, this question is the same as our
original question of relative phase between two quantum states).
So, how do we measure the angle when the two vectors are at
different positions? Well, we can transport one of them to the
other one and then, when they are next to each other, the angle is
easily measured (in the usual way as the ``angular distance"
between the vectors). But, again, we don't want the transport
itself to introduce any additional angles - we'd like it to be as
``straight" as possible. The straightest possible path is known as
a geodesic and the corresponding evolution along this path is
known as the {\em parallel transport} (the usefulness of this
concept in physics cannot be overemphasized - see \cite{Frankel}).
To define a parallel transport, let's look at the infinitesimal
evolution, from $|\psi (s)\rangle$ to $|\psi (s+ds)\rangle$. If we
don't want there to be any twists and turns in the phase, even
infinitesimally, then the two states should be in phase. So, we
require that
\begin{equation}
\mbox{Arg} \{\langle \psi (s)|\psi (s+ds)\rangle\} = 0
\end{equation}
This is the same as asking that $\langle \psi (s)|\psi
(s+ds)\rangle$ is purely real, i.e.
\begin{equation}
\mbox{Im} \{\langle \psi (s)|\psi (s+ds)\rangle\} = \mbox{Im}
\{\langle \psi (s)|d|\psi (s)\rangle\} = 0
\end{equation}
up to second order. But $\langle \psi (s)|d|\psi (s)\rangle$ is
purely imaginary anyway (prove it by differentiating $\langle \psi
(s)|\psi (s)\rangle=1$ by $ds$), and hence the parallel transport
condition becomes:
\begin{equation}
\langle \psi (s)|d|\psi (s)\rangle = 0 \label{pt}
\end{equation}
If the evolution satisfies this equation, then the phase is said
to be parallely transported, which is what we mentioned was
required in order to be able to define a relative phase between
two states (also see \cite{Anandan} for an accessible treatment of
the phase parallel transport) . This definition of parallel
transport is, however, {\em not} automatically gauge invariant. By
this we mean that if instead of the state $|\psi (s)\rangle$, we
use the state $|\tilde{\psi} (s)\rangle = e^{i\alpha (s)} |\psi
(s)\rangle$, then the parallel transport condition changes by the
amount
\begin{equation}
\langle \tilde{\psi} (s)|d|\tilde{\psi} (s)\rangle = \langle \psi
(s)|d|\psi (s)\rangle + i\frac{d\alpha}{ds} ds
\end{equation}
as can easily be checked. In order to obtain something that is
gauge invariant we can integrate the expression $\langle \psi
|d|\psi \rangle$ over a closed loop, giving us the expression for
the geometric phase, and then exponentiate the result (the
integral of $\frac{d\alpha}{ds}$ over a closed loop is $2\pi$,
whose exponential is equal to unity). So, the geometric phase
resulting from the parallel transport is
\begin{equation}
\gamma = \int_i^f \langle \psi (s)|\frac{d}{ds}|\psi (s)\rangle ds
\label{gp}
\end{equation}
and its exponential (over a closed loop) is gauge independent, but
{\em not} path independent \footnote{Path dependence cannot easily
be eliminated, although this is the key idea behind constructing
``topological" as opposed to ``geometric" phases}. Incidentally,
note that we are requiring that the phase difference between {\em
every} two infinitesimal points is zero. So, how come, if every
infinitesimal difference is zero, that on a finite stretch a
non-zero phase builds up? The answer is that the underlying space
is curved and it is the curvature that is reflected in the phase
difference; in fact, the curvature is equal the phase difference
(up to a constant factor). When a quantity vanishes
infinitesimally, but its integral over a finite region doesn't,
then this quantity is called {\em non-integrable}. Therefore, it
can be succinctly said that geometric phases are a manifestation
of non-integrable phase factors in quantum mechanics. Let's look
at a spin half system to illustrate this point.

\subsection{The Bloch Sphere}

Two level systems are ubiquitous in nature and can be used to
store and manipulate any form of information (in quantum computing
they are known as quantum bits or qubits). Two level systems are
very conveniently represented on a sphere, such that all pure
states lie on the surface of this sphere and all the mixed states
are inside. There is a one-to-one correspondence between the
points on and inside the sphere and quantum states of a two level
system (Fig 3). This is because we need at most three parameters
to specify every two level state, as can be seen from the general
density matrix for the systems:
\begin{equation}
\rho = \frac{1}{4} (I + \sum_i s_i \sigma_i)
\end{equation}
where $\sigma_i$ are the Pauli spin matrices and $s_i = \mbox{tr}
\sigma_i \rho$ are the $x,y,z$ coordinates on the sphere. If a
state is pure than $s^2_x+s^2_y+s^2_z = 1$ which, as we said, is a
point on the surface of the sphere (a pure state in other words).

\begin{figure}[ht]
\begin{center}
\hspace{0mm} \epsfxsize=7.0cm
\epsfbox{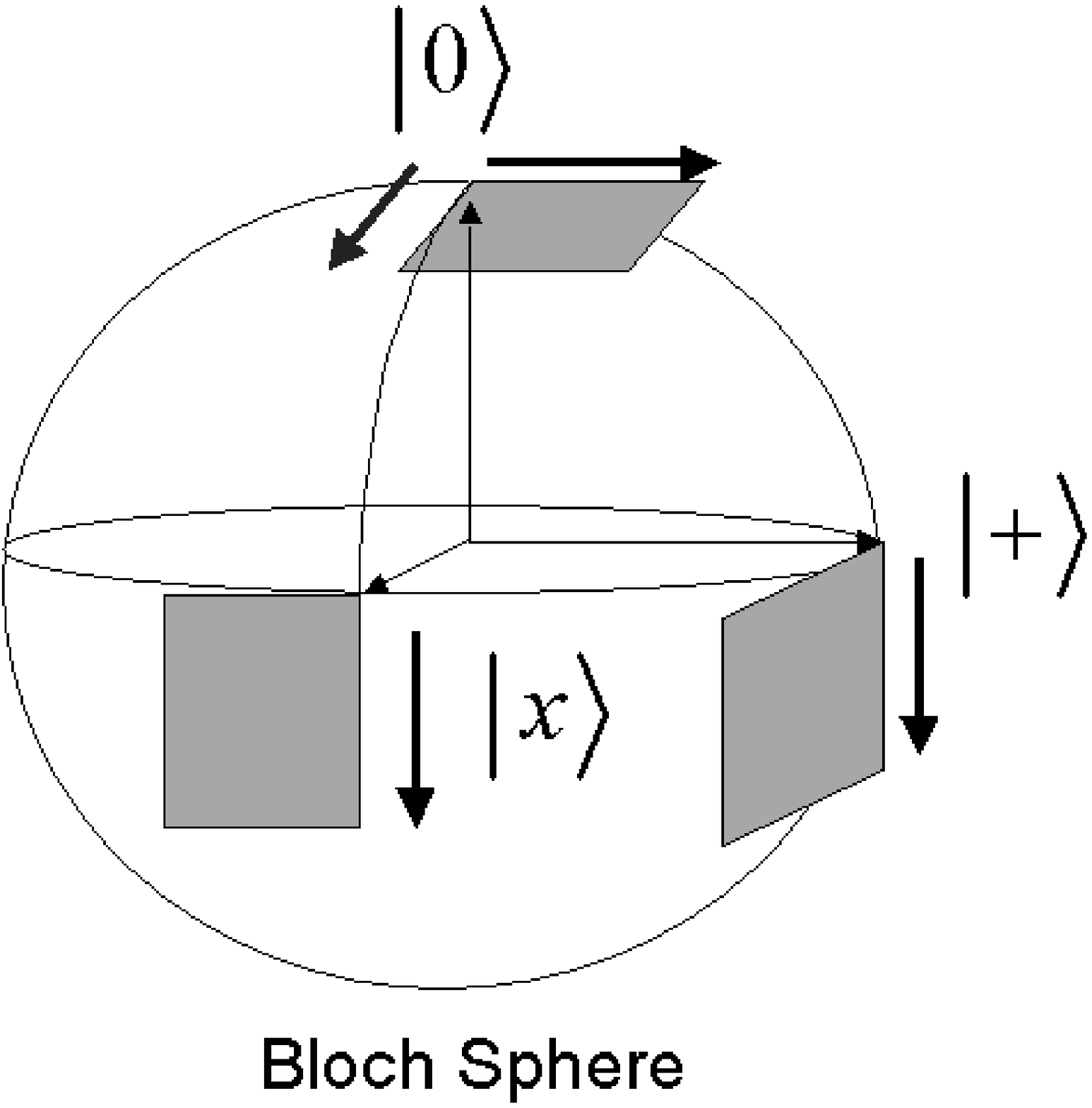}\\[0.2cm]
\begin{caption}
{Parallel transport of the phase on the Bloch sphere. The
presented evolution can be implemented in many different ways, one
of which is the adiabatic Schr\"odinger equation as explained in
the next section. The phase vector, initially at the North Pole
and lying in the plane of the paper, for example, is then
transported to the equator. After that, it moves on the equator,
and makes a $90$ degrees shift, before coming back to the North
Pole. The final phase vector, sticking out of the page, is at an
angle to the initial vector which is also $90$ degrees.}
\end{caption}
\end{center}
\end{figure}

Suppose that we now evolve from the state $|0\rangle$ to the state
$|+\rangle = |0\rangle + |1\rangle$, then to $|x\rangle =
|0\rangle + i|1\rangle$ and finally back to the state $|0\rangle$.
On the Bloch sphere we are going from the North Pole to the
Equator, then we move on the Equator by an angle of $\pi/2$ and
finally we move back to the North Pole). What is the corresponding
geometric phase? To see this we start with a tangential vector
initially at the North Pole pointing in some direction (there are
infinitely many directions corresponding to infinitely many
arbitrary phases to start with). If we now parallely transport
this vector along the described path, then we end up with a vector
pointing in a different direction to the original one (even
though, infinitesimally, the phase vector has always stayed
parallel to itself!). The angle between the initial and the final
vector is $\pi /2$, which is exactly equal to the area covered by
the state vector during the transport, or the corresponding solid
angle of the transport (the two are just different ways of
phrasing the same thing and can be seen to be equivalent using
Stokes' theorem). Therefore, the two orthogonal states $|0\rangle$
and $|1\rangle$ evolve in the following way:
\begin{eqnarray}
|0\rangle & \rightarrow & e^{i\Omega/2} |0\rangle\\
|1\rangle & \rightarrow & e^{-i\Omega/2} |1\rangle
\end{eqnarray}
where $\Omega$ is the solid angle (the factor of half is there
because the orthogonal states are $\pi$ away from each other in
the Bloch representation). This is important as it shows that
orthogonal states pick up opposite phases of equal magnitude. The
phase can be computed from the formula in eq. (\ref{gp}), but we
note that it can also be written in a ``discrete" way as
\begin{equation}
\arg \{\langle 0 |+\rangle\langle +|x\rangle\langle x|0\rangle \}
\end{equation}
This beautiful formula - originally due to Pancharatnam
\cite{Wilczek} - will be the basis of our general formulation of
the geometric phase for pure states. From it we will be able to
find that the space of quantum two level systems is curved. Are
there, for that matter, any quantum states whose space is not
curved (i.e. is flat) and how do we determine the curvature in
general? If you want to measure a curvature at a point, then make
a small loop around that point and compare the initial phase of
your state to the phase after completing this loop. Suppose that
the evolution is $|\psi (s_i)\rangle \rightarrow |\psi (s_i +
\delta s)\rangle \rightarrow |\psi (s_i + \delta s + \Delta
s)\rangle \rightarrow |\psi (s_i)\rangle$. Then, the final phase
difference is (according to Pancharatnam)
\begin{eqnarray}
\delta \theta & = & \arg \{ \langle \psi (s_i)|\psi (s_i + \delta
s)\rangle \langle \psi (s_i + \delta s)|\psi (s_i + \delta s +
\delta s')\rangle\langle \psi (s_i + \delta s + \delta
s')|\psi (s_i)\rangle \} \nonumber \\
& = & \arg \{ 1 + i (\frac{\delta}{\delta s} \langle \psi (s_i)|
\frac{\delta}{\delta s'} | \psi (s_i)\rangle -
\frac{\delta}{\delta
s'} \langle \psi (s_i)| \frac{\delta}{\delta s} | \psi (s_i)\rangle)\}\\
& = & \frac{\delta}{\delta s} \langle \psi (s_i)|
\frac{\delta}{\delta s'} | \psi (s_i)\rangle -
\frac{\delta}{\delta s'} \langle \psi (s_i) |\frac{\delta}{\delta
s} | \psi (s_i)\rangle
\end{eqnarray}
The definition of curvature, $K$, is $\delta \theta = K \delta A$,
where $A$ is the area enclosed (this definition of curvature is
due to Gauss). You can convince yourselves that for spin half
particles $\delta \theta = \delta A$, which means that the
curvature is $1$ (it is really $1/r^2$ as for any sphere, but the
Bloch sphere has a unit radius). For optical coherent states, on
the other hand, $\delta \theta = 0$ (if we vary the phase and the
amplitude of the state only), indicating that they line on a flat
surface (which you can prove as an exercise, but see
\cite{Provost} for more details on the Riemannian structure of
quantum states).

This formalism is very powerful and important in the modern
formulation of physical theories (the so called gauge theories -
see \cite{Frankel} for more details). But before we admire the
beauty and cleverness of all this any further, let's ask ourselves
a very basic question: how do we actually (physically) implement
the parallel transport given above?

\subsection{Adiabatic implementation of parallel transport}

Here we present one such implementation due to Berry \cite{Berry},
who actually discovered the quantum geometric phase in $1984$ in
this way. A wavefunction of a system is a function of a set of
parameters (which generally depend on time) and the time itself,
$|\Psi (s(t),t)\rangle $. Suppose that the Hamiltonian is a
function of $s$ only $H=H(s(t))$. Suppose, in addition, that these
parameters change very slowly so that the system, which starts in
a eigenstate of the initial Hamiltonian, stays an eigenstate of
the instantaneous Hamiltonian, {\em i.e.}
\[
H(s(t))|\Psi _{n}(s(t),t)\rangle =E_{n}(s(t))|\Psi_n
(s(t),t)\rangle
\]
(this is what is known as the adiabatic approximation in quantum
physics- the other extreme is the sudden approximation where the
Hamiltonian changes instantaneously). The Schr\"{o}dinger equation
for this system is
\[
i\frac{d}{dt}|\Psi_n (s(t),t)\rangle =H(s(t))|\Psi
_{n}(s(t),t)\rangle
\]
(where we assume throughout that $\hbar =1$). We would like to
show that the adiabatic evolution naturally implements the
parallel transport of the phase of the quantum state. Note that
apart from the adiabaticity assumption our analysis is completely
general. Multiplying the Schr\"odinger equation by $\langle
\Psi_n|$, and taking the eigenvalue equation into account we
obtain:
\[
i \langle \Psi_n | \frac{d}{dt} |\Psi_n \rangle = E_n
\]
Now, every state gains a dynamical phase (as well as a geometric
one) as it evolves. We'd like to get rid of the dynamical
component of the phase and not take it into account when we are
measuring the geometric contribution. To do so we define a new
wavefunction by taking into account the dynamical phase
\[
|\Phi _{n}(s(t),t)\rangle :=\exp \{iE_{n}(s(t))\}|\Psi
_{n}(s(t),t)\rangle
\]
which satisfies the following equation:
\[
\langle \Phi (s(t),t)|\frac{d}{dt}|\Phi (s(t),t)\rangle =0
\]
And this is the same as the parallel transport condition. So, the
geometric part of the quantum phase is parallely transported when
the state evolves under the Schr\"odinger equation in the
adiabatic approximation. So can we derive a closed formula for the
geometric phase? The answer is yes, and the solution to this
equation is given by:
\[
|\Phi (s(t),t)\rangle =e^{i\gamma (t)}|\Phi (s(t))\rangle
\]
where
\[
\frac{d}{dt}\gamma (t)=-i\langle \Phi (s(t))|\frac{d}{ds}|\Phi
(s(t))\rangle \frac{ds}{dt}
\]
We notice that the right hand side involves the 1-form $\beta $
describing the (symplectic) manifold of the projective Hilbert
space \footnote{Differential forms are just things that exist
under the integral sign, however, in mathematics they can be given
a life of their own. We will not need this formalization in the
rest of this review, but they feature very strongly in modern
theoretical physics, which is why I mention them here.} - see
Simon's original differential geometric interpretation
\cite{Simon}. Therefore,
\[
\frac{d}{ds}\gamma =\beta
\]
where $\beta = -i\langle \Phi (s(t))|\frac{d}{ds}|\Phi
(s(t))\rangle$. Integrating over a closed circuit $\partial S$ we
solve for the Berry phase
\[
\gamma =\oint_{\partial S}{\bf \beta }
\]
Invoking Stokes' theorem this becomes
\[
\alpha =\int_{S}{\bf \sigma }
\]
where $S$ is the area enclosed by the circuit $\partial S$ and
${\bf \sigma } $ is the 2-form on the symplectic manifold of the
projective Hilbert space. This phase is called the Berry phase
(the Berry and geometric phase are terms that I will use
interchangeably, although this is by no means the standard
convention in the literature).

In summary, the quantum phase arising in a general evolution of a
quantum system consists of two parts: the dynamical part given by
the formula
\[
\delta = \int E(t) dt
\]
and the geometric part, given by
\[
\gamma = \int \langle \Psi (s) | \frac{d}{ds} |\Psi (s)\rangle
\frac{ds}{dt} dt
\]
The total phase is then the sum of the two contributions, $\phi =
\delta + \gamma$.

We can encapsulate this whole subsection by saying that

{\em The Schr\"odinger equation in the adiabatic limit implements
the parallel transport of the phase of the evolving state.}

In order to become more acquainted with the notion of geometric
phases, we look at the classical analogue of this phenomenon.

\subsection{Classical geometric phase}

In this subsection we will talk about astronauts, cats and the
Foucault pendulum in this section, all of which are classical
objects. What do all these have in common? They all have the
geometric phase at the root of their basic behaviour! Suppose an
astronaut is in free space and initially facing away from his
space ship. He wants to turn around by $180$ degrees to return to
the ship, but there is nothing around him to push against or hold
onto in order to turn around. His task may at first sight look
impossible to achieve: his angular momentum is initially zero and
in order to turn he {\em seems to need} to generate some angular
momentum. But there are no forces externally, so this looks indeed
impossible! However, anyone who has been observing cats knows that
this reasoning is false. Cats face the same problem everyday and
they manage to solve it. If a cat falls from a certain height and
was initially upside down, she is still going to (almost always)
land on her feet (hence the allegorical expression: ``cats always
land on their feet"). So, how do cats manage to apparently
``violate the conservation of angular momentum"? Well, they don't
violate angular momentum conservation: they don't use dynamics to
turn, they use topology! Fig. 4 shows how the astronaut does
something similar to the cat to turn around. Note that his hand
movement is the same as the spin half evolution on the Bloch
sphere - hence the angle by which he turns can be calculated in
the same way.

\begin{figure}[ht]
\begin{center}
\hspace{0mm} \epsfxsize=10.0cm
\epsfbox{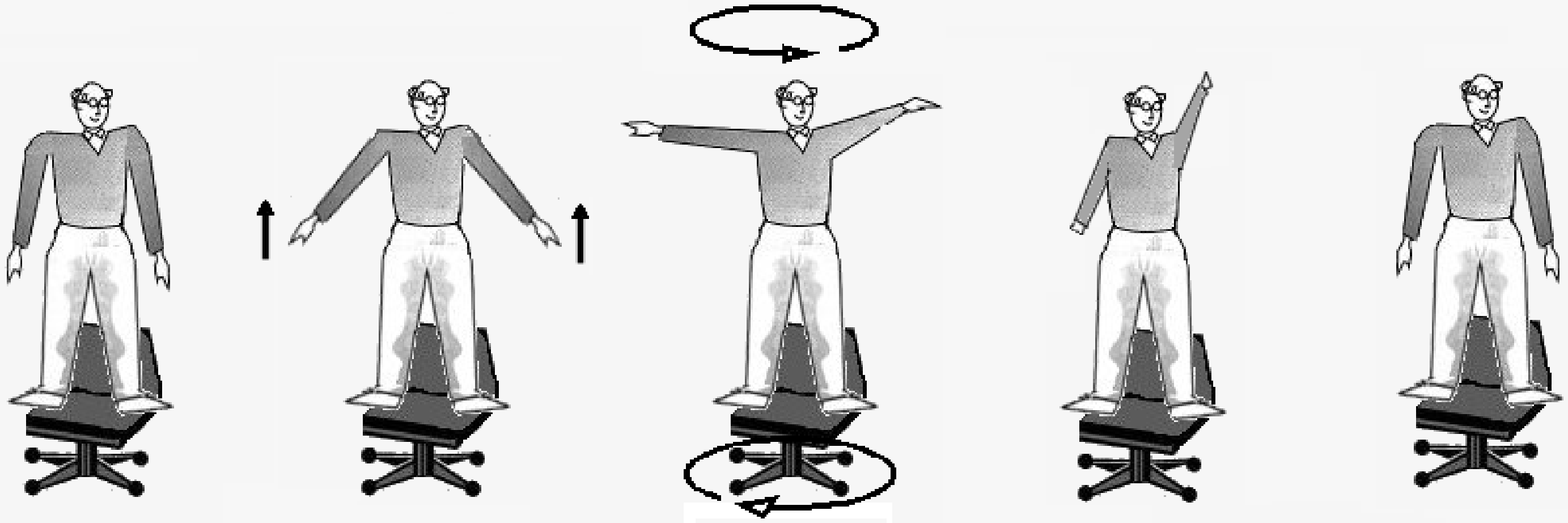}\\[0.2cm]
\begin{caption}
{A person rotating while standing on the chair by using a purely
geometrical effect while maintaining a zero angular momentum. In
the position three, the person turns his arms around, and at the
same time his body has to turn in the opposite direction to
maintain zero angular momentum on the whole. Therefore by
repeating the whole cycle a number of times, the person can make a
rotation on the chair by any desired angle (try it for yourself!).
This same effect is used by a cat to turn around in the air and
land on the feel, even when initially being upside down.}
\end{caption}
\end{center}
\end{figure}

Interestingly enough, the Foucault pendulum can be explained in
the same geometric way \cite{Hammond}. Imagine that a pendulum is
suspended at some latitude $\theta$, and that we observe its
swinging motion as the Earth spins around its own axis. It is well
known that the pendulum will after one rotation of the Earth swing
in a plane at an angle to the original plane of motion. To see
this, let us obtain an equation of motion of the pendulum. The
Lagrangian for this problem is given by
\begin{equation}
L = \frac{1}{2} m (\frac{dx}{dt}^2 + \frac{dy}{dt}^2 ) -
\frac{1}{2} m\omega^2 (x^2 + y^2) - m\Omega \cos \theta
(x\frac{dy}{dt} - y\frac{dx}{dt})
\end{equation}
where $\Omega$ is the Earth's frequency of rotation ($2\pi$ per
day), $m$ is the mass of the pendulum and $\omega$ its natural
frequency of swinging. The corresponding Euler-Lagrange equations
of motion are easy to obtain (exercise!) and their solution
written in the coordinate $z = x + iy$ is (in the adiabatic
limit):
\begin{equation}
z(t) \approx x_0 e^{-i\Omega \cos \theta t} e^{-i\omega t}
\end{equation}
This form of the solution is convenient as it allows us to see two
phases in the motion of the pendulum: the dynamical phase $\omega
t$ which is the same as the quantum dynamical phase seen before
and the geometric phase $\Omega \cos \theta t$. Note that after
one day revolution of the earth, the geometric phase is $2 \pi (1-
\cos \theta)$ which is the same as the spin half geometrical
phase. This is because our planet is a very good, albeit somewhat
imperfect, physical representation of the Bloch sphere!

\subsection{Experimental realization: The Pancharatnam connection}

I now summarize the theory so far: we would like to compare phases
of a quantum state at two different points. To do so we have to
chose a path connecting the two points, and evolve one of the
points along the path until it coincides with the other one. Then
we interfere them to infer the phase. But how does this
interference work?

Consider a conventional Mach-Zehnder interferometer in which the
beam-pair spans a two dimensional Hilbert space $\tilde{{\cal H}}
= \{ |\tilde{0}\rangle , |\tilde{1}\rangle \}$ \cite{Sjoqvist}.
The state vectors $|\tilde{0}\rangle$ and $|\tilde{1}\rangle$ can
be taken as (orthogonal) wave packets that move in two given
directions defined by the geometry of the interferometer. In this
basis, we may represent mirrors, beam-splitters and relative
$U(1)$ phase shifts by the unitary operators
\begin{eqnarray}
\tilde{U}_{M} & = & \left(
\begin{array}{rr} 0 & 1 \\
 1 & 0 \end{array} \right) , \, \,
\tilde{U}_{B} = \frac{1}{\sqrt{2}} \left(
\begin{array}{rr} 1 & 1 \\
 1 & -1 \end{array} \right) ,
\nonumber \\
\tilde{U}(1) & = & \left(
\begin{array}{cr} e^{i\chi} & 0 \\
 0 & 1 \end{array} \right) ,
\label{eq:spatialunitary}
\end{eqnarray}
respectively. An input pure state $\tilde{\rho}_{in} = | \tilde{0}
\rangle \langle \tilde{0} |$ of the interferometer transforms into
the output state
\begin{eqnarray}
\tilde{\rho}_{out} & = & \tilde{U}_{B} \tilde{U}_{M} \tilde{U}(1)
\tilde{U}_{B} \tilde{\rho}_{in} \tilde{U}_{B}^{\dagger}
\tilde{U}^{\dagger}(1) \tilde{U}_{M}^{\dagger}
\tilde{U}_{B}^{\dagger} \nonumber \\
 & = & \frac{1}{2} \left( \begin{array}{cc}
1 + \cos \chi & i \sin \chi  \\
-i \sin \chi & 1 - \cos \chi  \end{array} \right)
\end{eqnarray}
that yields the intensity along $|\tilde{0}\rangle$ as $I \propto
1+\cos \chi$. Thus the relative $U(1)$ phase $\chi$ could be
observed in the output signal of the interferometer.

Now assume that the particles carry additional internal degrees of
freedom, e.g., spin (it is this degree of freedom that will be
parallely transported). This internal spin space ${\cal H}_{i}
\cong {\cal C}^N$ is spanned by the vectors $|k\rangle$, $k=1,2,
\ldots N$. The density operator could be made to change inside the
interferometer
\begin{equation}
\rho_{0} \longrightarrow  U_{i} \rho_{0} U_{i}^{\dagger}
\end{equation}
with $U_{i}$ a unitary transformation acting only on the internal
degrees of freedom (we will see later on that this transformation
need not be unitary, but could be a more general completely
positive map). Mirrors and beam-splitters are assumed to leave the
internal state unchanged so that we may replace $\tilde{U}_{M}$
and $\tilde{U}_{B}$ by ${\bf U}_{M} = \tilde{U}_{M}\otimes 1_{i}$
and ${\bf U}_{B} = \tilde{U}_{B}\otimes 1_{i}$, respectively,
$1_{i}$ being the internal unit operator. Furthermore, we
introduce the unitary transformation
\begin{equation}
{\bf U} = \left( \begin{array}{rr}
 0 & 0 \\ 0 & 1 \end{array} \right) \otimes U_{i} +
\left( \begin{array}{cr}
 e^{i\chi} & 0 \\ 0 & 0 \end{array} \right) \otimes
1_{i} . \label{eq:unitary}
\end{equation}
The operators ${\bf U}_{M}$, ${\bf U}_{B}$, and ${\bf U}$ act on
the full Hilbert space $\tilde{{\cal H}} \otimes {\cal H}_{i}$.
${\bf U}$ corresponds to the application of $U_{i}$ along the
$|\tilde{1}\rangle$ path and the $U(1)$ phase $\chi$ similarly
along $|\tilde{0}\rangle$. We shall use ${\bf U}$ to generalize
the notion of phase to unitarily evolving mixed states.

\begin{figure}[ht]
\begin{center}
\hspace{0mm} \epsfxsize=7.0cm
\epsfbox{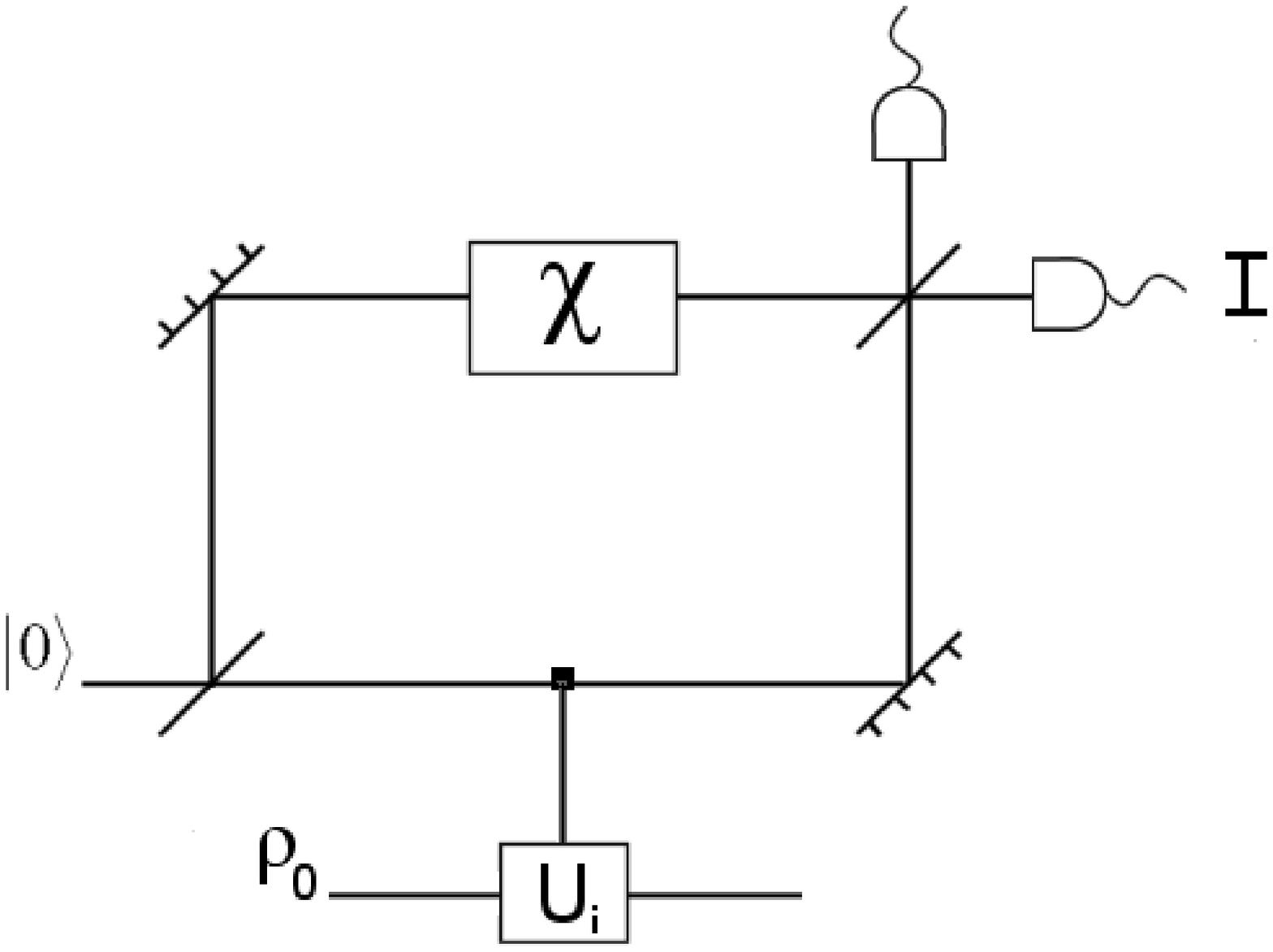}\\[0.2cm]
\begin{caption}
{The Mach Zehnder interferometer intended to implement the
measurement of the Pancharatnam phase. The internal degree of the
system going through the interferometer is manipulated only in one
branch of the interferometer. The manipulated branch is then
interfered at the end with the non-manipulated branch to infer the
phase difference.}
\end{caption}
\end{center}
\end{figure}

Let an incoming state given by the density matrix $\varrho_{in} =
\tilde{\rho}_{in} \otimes \rho_{0} = |\tilde{0} \rangle \langle
\tilde{0} | \otimes \rho_{0}$ be split coherently by a
beam-splitter and re-combine at a second beam-splitter after being
reflected by two mirrors. Suppose that ${\bf U}$ is applied
between the first beam-splitter and the mirror pair. The incoming
state transforms into the output state
\begin{equation}
\varrho_{out} = {\bf U}_{B} {\bf U}_{M} {\bf U} {\bf U}_{B}
\varrho_{in} {\bf U}_{B}^{\dagger} {\bf U}^{\dagger} {\bf
U}_{M}^{\dagger} {\bf U}_{B}^{\dagger} . \label{eq:outputmatrix}
\end{equation}
Inserting Eqs.~(\ref{eq:spatialunitary}) and (\ref{eq:unitary})
into Eq.~(\ref{eq:outputmatrix}) yields
\begin{eqnarray}
\varrho_{out} & = & \frac{1}{4} \left[
\left( \begin{array}{rr} \phantom{,}1 & \phantom{-}1 \\
1 & 1 \end{array} \right) \otimes U_{i} \rho_{0} U_{i}^{\dagger} +
\left(
\begin{array}{rr} 1 & -1 \\ -1 & 1
\end{array} \right) \otimes \rho_{0} \right.
\nonumber \\ & & + e^{i\chi} \left( \begin{array}{rr} 1 & 1 \\ -1
& -1 \end{array} \right) \otimes \rho_{0} U_{i}^{\dagger}
\nonumber \\ & & \left. +
e^{-i\chi} \left( \begin{array}{rr} \phantom{,}1 & -1 \\
1 & -1 \end{array} \right) \otimes U_{i} \rho_{0} \right] .
\end{eqnarray}
The output intensity along $|\tilde{0}\rangle$ is
\begin{eqnarray}
I & \propto & Tr \left( U_{i} \rho_{0} U_{i}^{\dagger} + \rho_{0}
+ e^{-i\chi} U_{i} \rho_{0} + e^{i\chi} \rho_{0} U_{i}^{\dagger}
\right)
\nonumber \\
 & \propto & 1 + |Tr \left( U_{i}
\rho_{0} \right)| \cos \left[ \chi - \arg Tr \left( U_{i} \rho_{0}
\right) \right] , \label{eq:outputintensity}
\end{eqnarray}
where we have used $Tr ( \rho_{0} U_{i}^{\dagger} ) = \left[ Tr
\left( U_{i} \rho_{0} \right) \right]^{\ast}$.

The important observation from Eq.~(\ref{eq:outputintensity}) is
that the interference oscillations produced by the variable $U(1)$
phase $\chi$ {\em is shifted by $\phi = \arg Tr \left( U_{i}
\rho_{0} \right)$ for any internal input state $\rho_{0}$,} be it
mixed or pure. This phase difference reduces for pure states
$\rho_{0} = |\psi_{0} \rangle \langle \psi_{0}|$ to the
Pancharatnam phase difference between $ U_{i} |\psi_{0} \rangle$
and $|\psi_{0} \rangle$. Moreover the visibility of the
interference pattern is $\nu = |Tr \left( U_{i} \rho_{0} \right)|
\geq 0$, which reduces to the expected $\nu = |\langle \psi_{0}
|U_{i}|\psi_{0} \rangle|$ for pure states.

It is clear from this experiment what it means to have a parallel
transport of phase, i.e. no phase introduced during an
infinitesimal evolution. We require that
\begin{equation}
\arg Tr \left( U (t,t+dt) \rho_{0} \right) = 0
\end{equation}
which is the same as the parallel transport condition presented
before in eq. (\ref{pt}). Suppose for that matter that we have no
dynamics and only projective transformations (the transformation
$U$ is in this case no longer unitary, but the same above analysis
still holds). Then $U = |\psi_n\rangle\langle \psi_{n-1}
|...|\psi_2\rangle\langle \psi_1|$ so that the formula for the
phase now becomes
\begin{equation}
\gamma = arg \{ \langle \psi_1|\psi_2\rangle \langle
\psi_2|\psi_3\rangle \ldots \langle \psi_n|\psi_1\rangle \}
\end{equation}
which is known as the Pancharatnam phase. Now, this is the most
general form of the phase for pure state, which leaves the motion
between the states completely arbitrary: it could be adiabatic,
and cyclic, but it could also be non-unitary and open. There are
many advantages of expressing the phase in this way:
\begin{itemize}
\item it is aesthetically pleasing!
\item it is manifestly gauge invariant since whenever we have the
ket $|\psi_j\rangle$ in the expansion, we also have its bra
present, $\langle\psi_j|$, meaning that all the extra phases of
individual states cancel. Thus, states only appear as projectors
which is the key to achieving independence of phase. As a slight
digression, those of you who know a bit about the lattice gauge
theory (see \cite{Muenster} for an excellent succinct
introduction) will immediately notice the similarity with Wilson's
loop construction. In order to obtain the action leading to the
geometric phase we have to sum up the infinitesimal loops, $W_P$,
over all points $P$. Each loop is defined as
\begin{equation}
W_P = \langle \psi (x_P)|\psi (x_P + dx) \rangle ...\langle \psi
(x_P+ dx + \delta x)|\psi (x_P) \rangle
\end{equation}
After a simple manipulation and using the Baker-Hausdorff formula
\[
\exp (\lambda A) \exp (\lambda B) = \exp (\lambda (A+B) +
\lambda^2/2[A,B] + O(\lambda^3),
\]
it can be shown that
\begin{equation}
W_P = e^{iF_{\mu \nu}dx_{\mu}\delta x_{\nu}}
\end{equation}
where the fictitious geometric field strength is
\begin{equation}
F_{\mu \nu} = \partial_{\mu} A_{\nu} - \partial_{\nu} A_{\mu} - i
[A_{mu},A_{nu}]
\end{equation}
and the geometric potential (also called connection in the
lectures) is
\begin{equation}
A_{\nu} = \langle \psi (x) |\partial_{\nu} |\psi (x) \rangle
\end{equation}
To obtain the Wilson loop over a finite range we just need to add
up all the infinitesimal loops such that $W = \sum_P W_P$. Pushing
the analogy with gauge theories we can think of a fictitious
topological charge $Q^T$ defined via Gauss' law:
\begin{equation}
\frac{1}{A} \oint B_{\mu \nu} dA = Q^T
\end{equation}
Also, a covariant derivative of a field $\Phi$ is given by
\begin{equation}
\Phi_{; \alpha} = \partial_{\alpha} \Phi + i A_{\alpha} \Phi
\end{equation}
and the field satisfies the following well known identity
\begin{equation}
F_{\mu \nu ; \rho} + F_{\nu \rho ; \mu} + F_{ \nu \rho ; \mu} = 0
\end{equation}
\item the formulation is very general. Any most general quantum evolution
(the so called Completely positive, trace preserving map) going
through the states above will lead to the same phase. There are
many issues in this generalization that are still unclear, but I
will not make any comments on this as it would obscure the
exposition (see Sj{\"o}qvist et al. \cite{Sjoqvist} for details).
\item offers us a way of generalizing the phase to mixed states
(as averaging over the corresponding pure states).
\end{itemize}

So, the geometric phase is independent of dynamics and gauge-free,
however, it does depend on the underlying geometry of the
evolution. To what extent can we eliminate the sensitivity to
geometry?

\subsection{Topology: The Aharonov-Bohm effect}

The Aharonov-Bohm effect \cite{AB} can be revisited and explained
in terms of the geometrical phase notion. Whereas the curvature in
the above examples was in some sense fictitious (i.e. not
generated by a ``real" field, but by the intrinsic curvature in
the geometry of quantum states), the Aharanov-Bohm effect is a
manifestation of the electro-magnetic field. From this perspective
we can even say that

{\em electro-magnetism is the gauge invariant manifestation of a
non-integrable quantum phase factor}.

In other words, this means that the electromegnetic field (more
precisely the potential) can be derived by postulating that the
Schr\"odinger equation is invariant under local phase change of
the wavefunction.

The Aharonov-Bohm set-up is the best example that explores the
effect of the electro-magnetic field on phase through a beam of
electrons in a double connected region where the field strength is
zero as in Fig. 6. The phase gained by the beam is
\begin{equation}
\exp \{\frac{-ie\Phi}{\hbar c}\}
\end{equation}
where $\Phi$ is the flux of the field. This phase has been
experimentally observed. We see that the phase is path dependent
(which is another way of saying that it is non-integrable). Even
though it is path dependent, if we make one round trip about the
flux, then we always gain the same amount of phase no matter what
the path is. This a nice topological feature of the Aharonov-Bohm
phase; the phase is then dependent on the number of times the flux
has been encircled (the so called {\em winding number} in
topology). It is this topological feature that we would like to
have when we implement quantum computation (this idea was first
presented in a seminal paper by Kitaev \cite{Kitaev}). However, we
would like the topology to be of a non-abelian (i.e.
non-commutative) character as we explain later on in the review.

\begin{figure}[ht]
\begin{center}
\hspace{0mm} \epsfxsize=7.0cm
\epsfbox{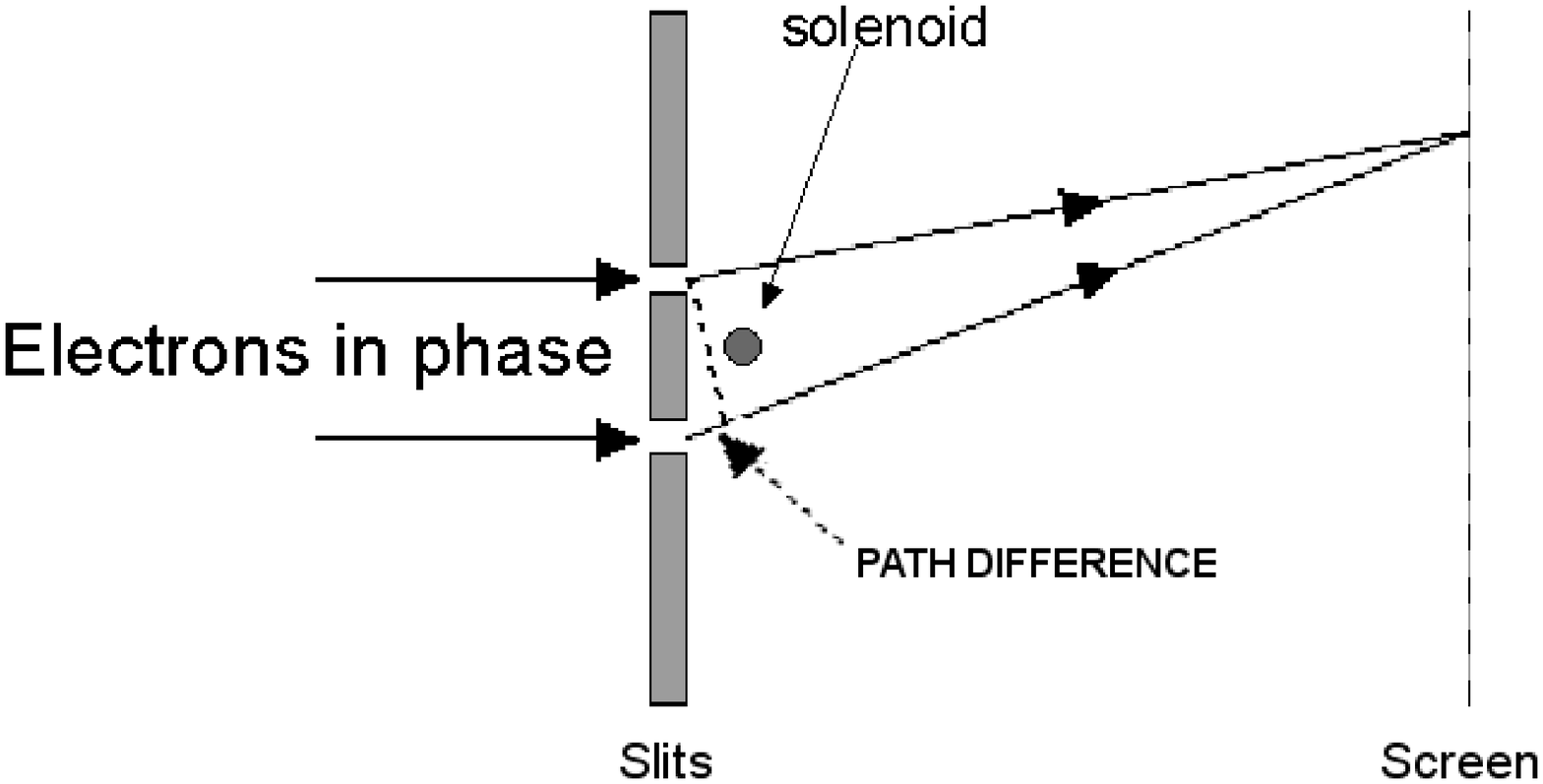}\\[0.2cm]
\begin{caption}
{The Aharonov Bohm set up. A beam of electrons undergoes a double
slit experiment with a solenoid placed behind the slits. The
soleniod generates a magnetic field which is confined to the small
region of the solenoid and is responsible for a shift in the
interference pattern at a screen placed some large distance away.}
\end{caption}
\end{center}
\end{figure}

\section{Quantum Computation}

So far we have explored connections between physics and geometry
through the concept of the geometric phase which we saw is
manifested both in quantum and classical mechanics. Now we plunge
into another equally exciting and cross-disciplinary area: quantum
information processing (see \cite{Nielsen} for details).

All information is ultimately encoded in the most elementary
quantum systems, so called quantum bits. A \emph{qubit} is a
quantum system in which the Boolean states $0$ and $1$ are
represented by a prescribed pair of normalised and mutually
orthogonal quantum states labeled as $\{|0\rangle ,|1\rangle \}$.
Unlike a simple Boolean variable, a qubit, typically a microscopic
system such as an atom, a nuclear spin, or a polarised photon, can
exist in an arbitrary superposition $\alpha |0\rangle + \beta
|1\rangle$, making it more powerful as a computational resource.

In quantum computation, we set some \emph{register} of qubits to
an ``input'' state, evolve the qubits unitarily using simple
building-block operations and then take the final state as
``output''. More formally, a \emph{quantum logic gate} is a device
which performs a fixed unitary operation on selected qubits in a
fixed period of time and a \emph{quantum network} is a device
consisting of quantum logic gates whose computational steps are
synchronised in time. The outputs of some of the gates are
connected by wires to the inputs of others. Also we define the
\emph{size} of the network to be the number of gates it contains.

\subsection{Quantum logic gates and universality}

The most common quantum gate is the Hadamard gate, a single qubit
gate $H$ performing the unitary transformation known as the
Hadamard transform.  It is defined as
\setlength{\unitlength}{0.030in}
\begin{equation}
H=\frac{1}{\sqrt{2}}\left(
\begin{array}{cc}
1 & 1 \\
1 & -1
\end{array}
\right) \mbox{\hspace{2cm}} \mbox{
\begin{picture}(30,0)(15,15)
  \put(-4,14){$|x\rangle$} \put(5,15){\line(1,0){5}} \put(20,15){\line(1,0){5}}
  \put(10,10){\framebox(10,10){$H$}}
\put(30,14){$\frac{(-1)^x\|x\rangle+|1-x\rangle}{\sqrt{2}}$}
\end{picture}
}
\end{equation}
The matrix is written in the computational basis $\{|0\rangle,
|1\rangle \}$ and the diagram on the right provides a schematic
representation of the gate $H$ acting on a qubit in state
$|x\rangle$, with $x=0,1$.

The addition of another single qubit gate, the phase shift gate
$\mathbf{\phi }$, defined as $\left| \,0\right\rangle \mapsto
\left| \,0\right\rangle $ and $\left| \,1\right\rangle \mapsto
e^{i\phi }\left| \,1\right\rangle $, or, in matrix notation,
\setlength{\unitlength}{0.030in}
\begin{equation}
{\mathbf{\phi}} = \left (
\begin{array}{cc}
1 & 0 \\
0 & e^{i\phi}
\end{array}
\right ) \mbox{\hspace{3cm}} \mbox{
\begin{picture}(30,0)(15,15)
  \put(-4,14){$|x\rangle$} \put(5,15){\line(1,0){20}} \put(20,15){\line(1,0){5}}
  \put(15,15){\circle*{3}}
\put(14,19){$\phi $} \put(30,14){$e^{ix\phi}|x\rangle$}
\end{picture}
}
\end{equation}
is actually sufficient to construct the following network (of size
four), which generates the most general pure state of a single
qubit (up to a global phase), \setlength{\unitlength}{0.025in}
\begin{equation}\label{1quniv}
\mbox{
\begin{picture}(80,12)(0,3)

\put(0,5){$|0\rangle$}

\put(10,5){\line(1,0){10}} \put(20,0){\framebox(10,10){H}}
\put(30,5){\line(1,0){20}} \put(50,0){\framebox(10,10){H}}
\put(60,5){\line(1,0){20}}

\put(40,5){\circle*{3}} \put(38,10){$2\theta$}

\put(70,5){\circle*{3}} \put(65,10){$\frac{\pi}{2}+\phi$}

\end{picture}
} \quad \cos\theta|0\rangle+e^{i\phi}\sin\theta|1\rangle.
\end{equation}
Consequently, the Hadamard and phase gates are sufficient to
construct \emph{any} unitary operation on a single qubit. We have
used both the above gates in our discussion of the Pancharatnam
phase in the Mach-Zehnder interferometer.

Thus the Hadamard gates and the phase gates can be used to
transform the input state $|0\rangle |0\rangle ...|0\rangle $ of
$n$ qubits into any state of the type $|\Psi _{1}\rangle $ $|\Psi
_{2}\rangle ...$ $ |\Psi _{n}\rangle ,$ where $|\Psi _{i}\rangle $
is an arbitrary superposition of $|0\rangle $ and $|1\rangle .$
These are rather special $n$-qubit states, called the product
states or the separable states. In general, a register of $n$
qubits can be prepared in states which are not separable, known as
entangled states.

However, in order to entangle two (or more qubits) it is necessary
to have access to two-qubit gates. One such gate is the controlled
phase shift gate $B(\phi)$ defined as
\begin{equation}
{B}(\phi )=\left. \left(
\begin{array}{cccc}
1 & 0 & 0 & 0 \\
0 & 1 & 0 & 0 \\
0 & 0 & 1 & 0 \\
0 & 0 & 0 & e^{i\phi }
\end{array}
\right) \mbox{\hspace{1.5cm}} \mbox{
\begin{picture}(25,0)(0,20)
  \put(-4,14){$|y\rangle$} \put(-4,29){$|x\rangle$} \put(5,15){\line(1,0){20}}
\put(5,30){\line(1,0){20}}
  \put(15,30){\circle*{3}} \put(15,15){\line(0,1){15}}
\put(15,15){\circle*{3}}
\end{picture}}\quad \right\} e^{ixy\phi }\left| \,x\right\rangle \left|
\,y\right\rangle .
\end{equation}
The matrix is written in the computational basis $\{ |00\rangle,
|01\rangle, |10\rangle,$ $|11\rangle \}$ and the diagram on the
right shows the structure of the gate. Note: we know how to
implement the single qubit phase gate geometrically, but not the
other two gates aforementioned. Which gates are important to be
able to implement in order to perform any quantum computation?

An important result in the theory of quantum computation states
that the Hadamard gate, and all $B(\phi)$ controlled phase gates
form an {\em universal set of gates}: if the Hadamard gate as well
as all $B(\phi)$ gates are available then any $n$-qubit unitary
operation can be simulated exactly with less than $C4^{n}n$ such
gates, for some constant $C$ \cite{Nielsen}.  Consequently, being
able to implement 1- and 2- qubit phase gates is of crucial
importance in quantum computation.  In the remaining part of the
lectures we describe a new method for implementing the controlled
phase gates based explicitly on geometric phases rather than
dynamic ones. Thus if we show that the universal set can be
implemented geometrically, then any computation can be implemented
geometrically (see \cite{Jones} for an NMR geometric
implementation of the controlled phase gate).

\subsection{Lightning review of quantum algorithms}

Deutsch's problem \cite{Deutsch} is the simplest possible example
which illustrates the advantages of quantum computation
\cite{Vedral}. The problem is the following. Suppose that we are
given a binary function of a binary variable $f: \{ 0,1\}
\longrightarrow \{ 0,1\}$. Thus, $f(0)$ can either be $0$ or $1$,
and $f(1)$ likewise can either be $0$ or $1$, giving altogether
four possibilities. However, suppose that we are not interested in
the particular values of the function at $0$ and $1$, but we need
to know
whether the function is: 1) constant, i.e. $f(0)=f(1)$, or 2) varying, i.e. $%
f(0)\ne f(1)$. Now Deutsch poses the following task: by computing
$f$ only {\em once} determine whether it is constant or varying.
This kind of problem is generally referred to as a {\em promise
algorithm}, because one property out of a certain number of
properties is initially promised to hold, and our task is to
determine computationally which one holds.

First of all, classically finding out in one step whether a
function is constant or varying is clearly impossible. We need to
compute $f(0)$ and then compute $f(1)$ in order to compare them.
There is no way out of this double evaluation. Quantum
mechanically, however, there is a simple method to achieve this
task by computing $f$ only once! We need to implement the
following unitary transformation
\begin{eqnarray}
|x\rangle  \Rightarrow e^{i\pi f(x)} |x\rangle \nonumber
\end{eqnarray}
where $x=0,1$. Now we can see where the power of quantum computers
is fully realised: {\em each of the components in the
superposition of $|0\rangle + |1\rangle$ undergoes the same above
evolution ``simultaneously"}, leading to the powerful ``quantum
parallelism". This feature is true for quantum computation in
general. Let us look at the two possibilities now:
\begin{enumerate}
\item  if $f$ is constant, then the final state is
\begin{eqnarray}
|\Psi _{\mbox{out}}\rangle =\pm (|0\rangle +|1\rangle )
\;.\nonumber
\end{eqnarray}
and
\item  if, on the other hand, $f$ is varying, then, the final state is
\begin{eqnarray}
|\Psi _{\mbox{out}}\rangle = \pm (|0\rangle
-|1\rangle)\;.\nonumber
\end{eqnarray}
\end{enumerate}
Note that the output qubit emerges in two different {\em
orthogonal} states, depending on the type of $f$. These two states
can be distinguished with $100$ percent efficiency. This is easy
to see if we first perform a Hadamard transformation on this
qubit, leading to the state $|0\rangle$ if the function is
constant, and to the state $|1\rangle$ if the function is varying.
Now a single projective measurement in $0,1$ basis determines the
type of the function. Therefore, unlike their classical
counterparts, quantum computers can solve Deutsch's problem.

It should be emphasised that this quantum computation, although
extremely simple, contains all the main features of successful
quantum algorithms: it can be shown that all quantum computations
are just more complicated variations of Deutsch's problem. In
particular, Shor's algorithm for factorisation of numbers is just
a scaled up version of Deutsch as it is based on period finding.
Note, finally, that the main operation in Deutsch's algorithm was
the introduction of a phase conditional on the state of the qubit.
This phase can, of course, be introduced in a purely geometrical
way, by a spin-half encircling an area of the size of $2\pi$.

\section{Topological implementation of quantum gates}

Speaking somewhat loosely, we would now like to construct a
universal set of quantum gates, which we know is impossible if we
only have the (abelian) geometric phases (because we cannot
implement a Hadamard transform with them). However, in addition to
achieving the universality, we'd like to be able to achieve a
certain degree of resistance to errors, but without performing any
error correction. How far can this idea be pushed?

\subsection{Non-abelian formalism}

The best way of explaining the non-abelian phases is to look at
how they arise during quantum evolution. The term ``non-abelian"
refers to the fact that two different phases don't commute. Note
that the ordinary (Berry) phases we discussed so far are abelian,
i.e.
\[
e^{i\gamma_1} e^{i\gamma_2} = e^{i\gamma_2}e^{i\gamma_1}
\]
This inequality, however, would not be satisfied if we had
non-abelian phases. This therefore implies that the non-abelian
phases would not be represented by ordinary numbers, but by
matrices. We now show that this scenario arises during the
Schr\"odinger evolution in the adiabatic setting when our system
is degenerate. Suppose that we have
\begin{eqnarray}
H (t) |\psi_1 (t) \rangle & = & E(t) |\psi_1 (t) \rangle \\
H (t) |\psi_1 (t) \rangle & = & E(t) |\psi_1 (t) \rangle
\end{eqnarray}
and $\langle \psi_1 (t)|\psi_2 (t)\rangle =0$. From the
Schr\"odinger equation we can conclude:
\begin{eqnarray}
i\langle \psi_1|\frac{d}{dt} |\psi \rangle & = & E \langle \psi_1 |\psi \rangle \\
i\langle \psi_2|\frac{d}{dt} |\psi \rangle & = & E \langle \psi_2
|\psi \rangle
\end{eqnarray}
We recast this pair of equations by writing $|\psi\rangle = a_1(t)
|\psi_1\rangle + a_2(t)|\psi_2\rangle$, and subtracting the
dynamical phases by putting $E(t) = 0$ (for all time $t$) we
obtain
\begin{eqnarray}
\frac{da_1}{dt} + a_1\langle \psi_1 |\frac{d}{dt}|\psi_1 \rangle +
a_2 \langle \psi_1|\frac{d}{dt} |\psi_2 \rangle & = & 0\\
\frac{da_2}{dt} + a_2\langle \psi_2 |\frac{d}{dt}|\psi_2 \rangle +
a_1 \langle \psi_2|\frac{d}{dt} |\psi_1 \rangle & = & 0
\end{eqnarray}
We can rewrite this equation as a vector-matrix equation,
$$
\frac{dv}{dt} = i \left(\begin{array}{cc} i\langle \psi_1
|\frac{d}{dt}|\psi_1 \rangle & i\langle \psi_1|\frac{d}{dt}
|\psi_2 \rangle \\ i\langle \psi_2|\frac{d}{dt} |\psi_1 \rangle &
i\langle \psi_2 |\frac{d}{dt}|\psi_2 \rangle \end{array}\right) v
$$
where
$$
v (t) = \left(\begin{array}{c} a_1 (t) \\ a_2 (t)
\end{array}\right)
$$
The formal solution to this equation can be written in the usual
``Dyson perturbative expansion":
\[
v(s) = {\cal P} \exp \{i\int_i^f A^j (s) T_j ds \}v(0)
\]
where ${\cal P}$ indicated the path-ordering operation of
non-commuting factors $A^j (s) T_j$. This non-abelian phases is in
general very difficult to evaluate, because of the path ordering
operation. Note that if the state starts in the degenerate
subspace it always remains there according to the adiabatic
theorem \footnote{We mention that the $SU(2)$ Yang - Mills
formalism is based on the non-abelian gauge invariance and can
reproduce the weak interaction and the $SU(3)$ formalism due to
Gell-Mann can be used to represent the strong interaction. These
are just generalizations of the $U(1)$ formalism for
Elactro-magnetism based of Weyl's ideas. The interested reader is
advised to read \cite{Frankel}}.

I would like to mention potential advantages of using topological
evolution to implement quantum gates. First of all, there is no
dynamical phase in the evolution. This is because we are using
degenerate states to encode information so that the dynamical
phase is the same for both states (and it factors out as it were).
Also, all the errors stemming from the dynamical phase are
automatically eliminated. Secondly, the states being degenerate do
not suffer from any bit flip errors between the states (like the
spontaneous emission). So, the evolution is protected against
these errors as well. Thirdly, the size of the error depends on
the area covered and is therefore immune to random noise (at least
in the first order) in the driving of the evolution. This is
because the area is preserved under such a noise as formally
proven by Palma \cite{Palma}. Also, by tuning the parameters of
the driving field it may be possible to make the phase independent
of the area to a large extent and make it dependent only on a
singular topological feature - such as in the Aharonov-Bohm effect
where the flux can be confined to a small area - and this would
then make the phase resistant under very general errors.

So, in order to see how this works in practice we take an atomic
system as our model implementing the non-abelian evolution. We'll
see that quantum computation can easily be implemented in this
way. The question, of course, is the one about the ultimate
benefits of this implementation. Although there are some obvious
benefits, as listed above, there are also some serious
shortcomings, and so the jury is still out on this issue.

\subsection{Example}

Let's look at the following $4$ level system analyzed by Unanyan,
Shore and Bergmann \cite{Shore}. They considered a four level
system with three degenerate levels $1,3,4$ and one level $2$ with
a different energy as in Fig 7. This system stores one bit of
information in the levels $1$ and $2$ (hence there is double the
redundancy in the encoding of information). We have the following
Hamiltonian
$$
H(t) = \left(\begin{array}{cccc} 0 & P(t) & 0 & 0 \\
P(t) & 0 & S(t) & Q(t) \\ 0 & S(t) & 0 & 0 \\ 0 & Q(t) & 0 & 0
\end{array}\right)
$$
where $P,Q,S$ are some functions of time (to be chosen at
leisure). It is not difficult to find eigenvalues and eigenvectors
of this matrix (exercise!). There are two degenerate eigenvectors
(with the corresponding zero eigenvalue for all times) which will
be implementing our qubit and they are
$$
\Phi_1 (t) = \left(\begin{array}{c} \cos \theta (t) \\
0  \\ -\sin \theta (t)  \\ 0 \end{array}\right)
$$
and
$$
\Phi_2 (t) = \left(\begin{array}{c} \sin \phi (t) \sin \theta (t) \\
0  \\ \sin \phi (t) \cos \theta (t)  \\ -\cos \phi (t)
\end{array}\right)
$$
where $\tan \theta (t) = P(t)/Q(t)$ and $\tan \phi (t) =
Q(t)/\sqrt{P(t)^2+Q(t)^2}$.

\begin{figure}[ht]
\begin{center}
\hspace{0mm} \epsfxsize=7.0cm
\epsfbox{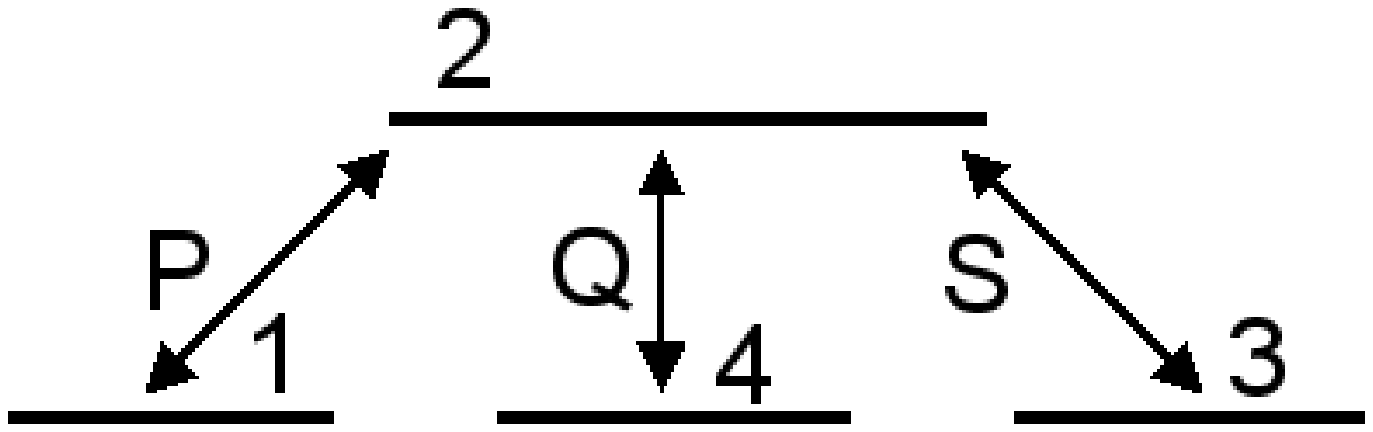}\\[0.2cm]
\begin{caption}
{The four level system that can be used for non-abelian quantum
computation to encode one qubit of information in two degenerate
levels. The method is detailed in the text.}
\end{caption}
\end{center}
\end{figure}

In the adiabatic limit (which we will assume here) these are the
only important ones as the state will remain in their subspace
throughout the evolution (this is guaranteed by the adiabatic
theorem). Although, in general, the Dyson equation is difficult to
solve, in this special example we can write down a closed form
expression \cite{Shore}. The unitary matrix representing the
geometrical evolution of the degenerate states is
$$
B(t) = \left(\begin{array}{cc} \cos \gamma (t) & \sin \gamma (t)
\\ -\sin \gamma (t) & \cos \gamma (t) \end{array}\right)
$$
where
\begin{equation}
\gamma (t) = \int \sin \phi (t) \frac{d\theta}{dt} dt
\end{equation}
This therefore allows us to calculate the non-abelian phase for
any closed path in the parametric space. After some time we
suppose that the parameters return to their original value. So, at
the end of the interaction we have the matrix
$$
B(\infty) = \left(\begin{array}{cc} \cos \gamma_f  & \sin \gamma_f
\\ -\sin \gamma_f & \cos \gamma_f  \end{array}\right)
$$
where
\begin{equation}
\gamma_f = \oint_c \frac{Q}{(P^2 + S^2)\sqrt{Q^2 + P^2 + S^2}}
(SdP - PdS)
\end{equation}
which can be evaluated using Stokes' theorem (the phase will in
general depend on the path, as explained before). So, we can have
a non-abelian phase implementing a Hadamard gate. With two systems
of this type (mutually interacting) we can implement a controlled
Not gate and therefore (at least in principle) have a universal
quantum computer (see \cite{Zanardi}). As an exercise I leave it
to the reader to implement Deutsch's algorithm in the purely
topological way.

\section{\noindent Conclusions and Outlook}

We have seen what geometric phases are and appreciated their
importance in physics in general. They were shown to be linked to
the modern ideas of gauge theories and differential geometry, in
particular the idea of parallel transport. In addition to their
fundamental value, they also have a potential to implement quantum
gates that are, at least in principle, more reliable than any
existing gates. In order to do so I have argued that a high degree
of topological dependence has to be achieved - i.e. independence
of global geometry, which is certainly not that easy in practice.
These topological gates will most likely have to be combined with
other error correcting mechanisms in order to make a reliable
quantum computer. The exact engineering design will, in the end,
depend on the actual medium that is chosen to support quantum
computation. It is fair to say that when this review was completed
(in December 2002) there was no fully fault tolerant
implementation of quantum computation. Fortunately there has been
a huge number of proposals to implement such computations with
various solid-state systems (such as Josephson Junctions and
Quantum Dots), Nuclear Magnetic Resonance, Trapped Atoms and so
on, and the whole field is, therefore, very much alive at present.

\noindent {\bf Acknowledgments. } I have benefited greatly from
discussions with J. Anandan, S. Bose, A. Carollo, A. Ekert, P.
Falci, S. Fazio, I. Fuentes-Guridi, J. Jones, D. Markham, J.
Pachos, M. Palma, M. Santos, J. Siewert and E. Sj\"{o}qvist on the
subject of geometric phases. My research on this subject is
supported by European Community, Engineering and Physical Research
Council in UK, Hewlett-Packard and Elsag spa.

\bigskip

\bigskip

\end{document}